\documentclass{webofc}
\usepackage[varg]{txfonts}   
\usepackage{xcolor}

\begin{document}

\title{Dense-matter equation of state at zero \& finite temperature}

\author{\firstname{Alexander} \lastname{Clevinger}\inst{1},
        \firstname{Veronica} \lastname{Dexheimer}\inst{1},
        \firstname{Jeffrey} \lastname{Peterson}\inst{1}
}

\institute{Department of Physics, Kent State University, Kent, OH 44243, USA}

\abstract{At high density, matter is expected to undergo a phase transition to deconfined quark matter. Although the density at which it happens and the strength of the transition are still largely unknown, we can model it to be in agreement with known experimental data and reliable theoretical results. We discuss how deconfinement in dense matter can be affected by both by temperature and by strong magnetic fields within the CMF model. To explore different dependencies in our approach, we also explore how deconfinement can be affected by the assumption of different degrees of freedom, different vector coupling terms, and different deconfining potentials, all at zero temperature. Both zero-net-strangeness and isospin-symmetric heavy-ion collision matter and beta-equilibrated charge-neutral matter in neutron stars are discussed.}
\maketitle
%

There are not many models available that can self-consistently address deconfinement to quark matter and chiral symmetry restoration in dense matter. Such approaches are essential when one wants to address how the phase transition (we assume that both are related) changes as a function of temperature. To do so, we make use of the Chiral Mean Field (CMF) model, which is a relativistic approach based on an SU(3) non-linear realization of the sigma model \cite{Papazoglou:1998vr}. It has been extended to contain hadrons (nucleons and hyperons) and quarks \cite{Dexheimer:2009hi}, and can describe both matter created in heavy-ion collisions and in neutron stars when imposing different constraints related to charge or isospin, strangeness, and leptons. The model is constrained by several properties of low-energy nuclear physics and astrophysics, as well as reliable theoretical predictions of e.g., lattice QCD (see Ref.~\cite{MUSES:2023hyz} for a complete list of constraints). In this work, we follow Ref.~\cite{Peterson:2023bmr} and explore how the phase transition changes as a function of temperature and magnetic field. To estimate the model-dependency of our results, we make comparisons at finite temperature and magnetic fields with another model that can self-consistently address deconfinement and chiral symmetry, the Polyakov-loop extended Nambu-Jona-Lasinio (PNJL) model \cite{Meisinger:1995ih,Fukushima:2003fw}. 

Additionally, to study dependencies within our model at zero temperature, we follow \cite{Clevinger:2022xzl}. We consider a different deconfining potential and discuss how accounting for different degrees of freedom (including now spin 3/2 $\Delta$'s) and different forms of vector coupling terms can affect the interior of neutron stars, including stability against gravitational collapse. In particular, changing the baryon chemical potential dependency in our (Polyakov inspired) potential by using a lower power of $\mu_B$, drastically decreases the strength of the phase transition, leading to a smaller jump in energy density $\varepsilon$. While scalar interactions are constrained to reproduce vacuum masses of baryons in chiral models, together with other vacuum properties, vector interactions are constrained to reproduce nuclear matter and astrophysics. In particular, new results from electromagnetic and gravitational observation of neutron stars point to a soft equation of state (EoS) at low density to reproduce small stars \cite{Riley:2019yda,Miller:2019cac,LIGOScientific:2018cki} and a stiffer EoS at large densities to reproduce massive stars \cite{Fonseca:2021wxt}. In this work, we focus on two different vector self-interactions, a higher-order term and a mixed isoscalar-isovector one.


To include magnetic-field effects, besides Landau quantization, we also include the anomalous magnetic moment of hadrons and leptons (not quarks) to account for how particles with opposite spin react to the magnetic field. First discussing neutron-star matter, Fig.~1 left panel shows how a stronger phase transition takes place at larger energy density $\varepsilon$ (corresponding to larger baryon chemical potential $\mu_B$) for larger magnetic field strength $B$. These changes are further enhanced by the AMM. At finite temperature, Fig.~1 right panel shows that a weaker phase transition takes place at lower $\mu_B$ for larger temperatures $T$. Note that, while the temperature behavior agrees with PNJL calculations from \cite{Ferreira:2013tba,Ferreira:2014kpa,Costa:2015bza}, this is not the case for the magnetic field considerations.
Even more interesting, changing constraints from the ones describing neutron-star matter to the ones describing heavy-ion collisions can also change considerably deconfinement to quark matter. As shown in the right panel of Fig.~1, the phase transition takes place at larger $\mu_B$ and is stronger for
heavy-ion collision matter (for any $T$ and $B$) in CMF model. This result is not reproduced in the PNJL model.

\begin{figure*}[t!]
\centering
\includegraphics[trim={1.3cm 1.cm 2cm 2cm},clip,width=0.496\linewidth]{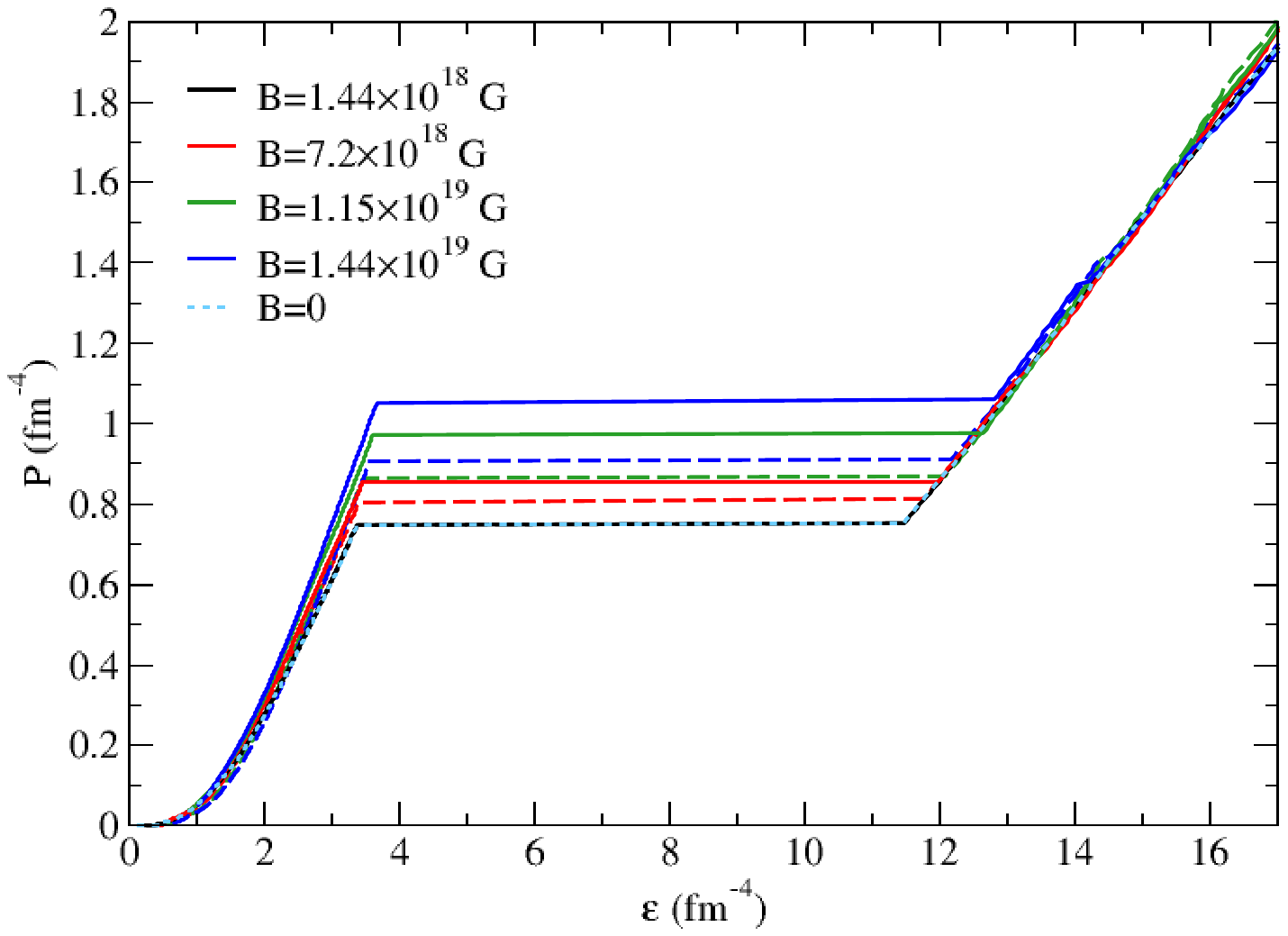}
\includegraphics[trim={1.3cm 1.cm 2cm 2cm},clip,width=0.496\linewidth]{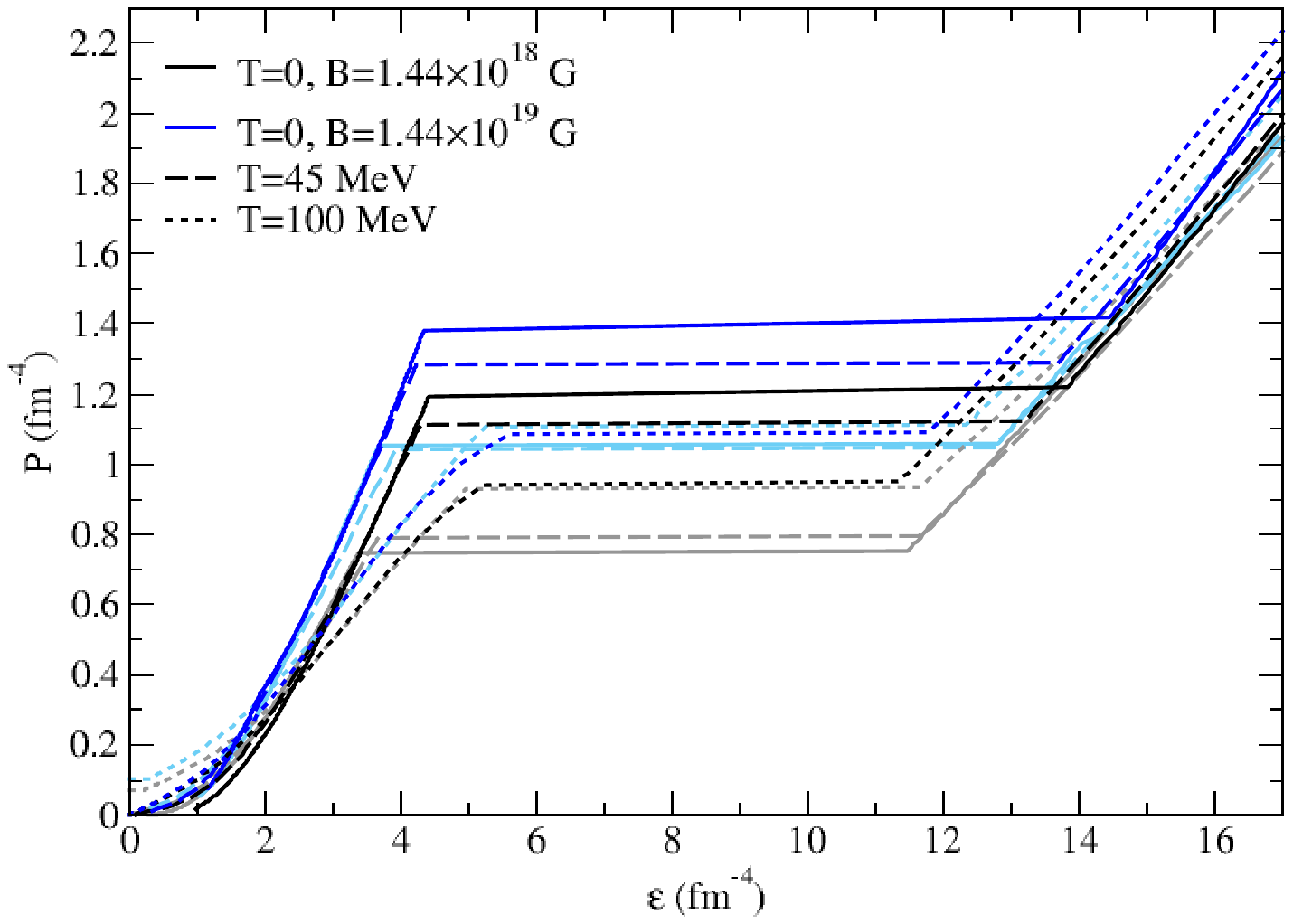}
\caption{Left: Pressure vs. energy density for neutron-star matter at $T=0$ for several magnetic-field strengths with (solid) and without (dashed) AMM. Right: EoS for neutron-star and heavy-ion-collision matter (gray instead of black and light blue instead of blue) for two temperatures and the strongest/weakest magnetic-field strengths with AMM.}
\label{eosT0Star}
\end{figure*}


\begin{figure*}[t!]
  \includegraphics[trim={0 .3cm 0 .1cm},clip,width=1.01\linewidth]{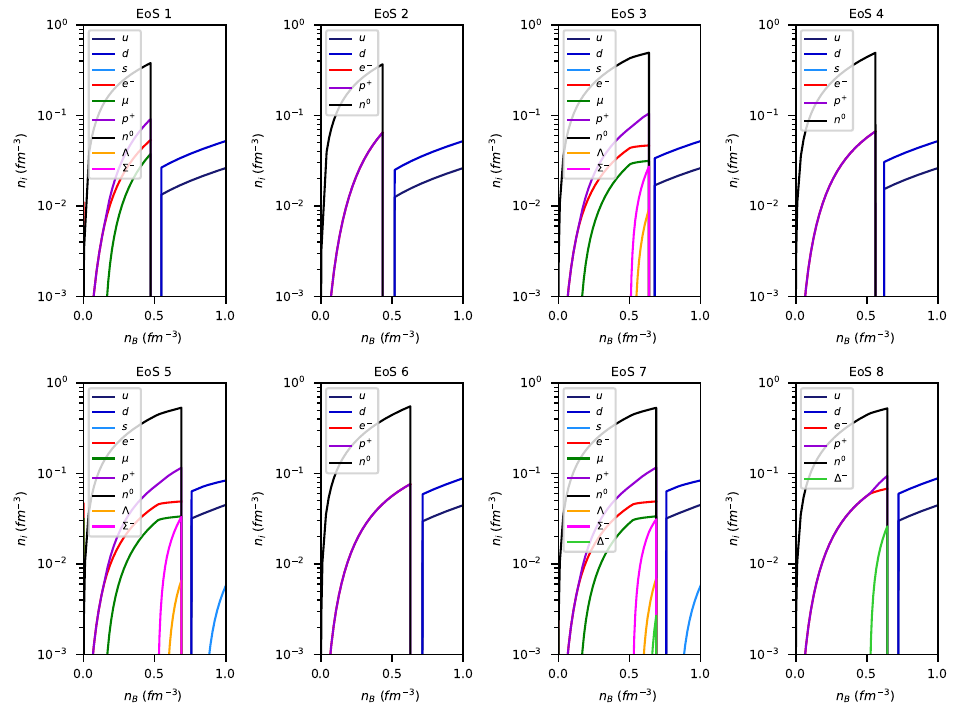}
  \caption{Particle population as a function of baryon density for different EoS's. For EoS's 2, 4, and 6, the line for the electrons coincides with the one for the protons. Quark densities are divided by $3$.}
  \label{fig:2}
\end{figure*}

In a new framework of the CMF model with a modified quark description, we study a weaker deconfinement phase transition at zero temperature \cite{Clevinger:2022xzl}. In Fig 2, we present 8 hybrid EoS with varying interaction terms and degrees of freedom. EoS 1, 3, 5, \& 7 include hyperons, muons and strange quarks while EoS 2, 4, 6, \& 8 do not (only nucleons and light quarks). EoS 7-8 also include $\Delta$ baryons. For EoS 1-2, we only include standard CMF terms while EoS 3-8 contain an additional $\omega\rho$ term and EoS 5-8 contain a higher-order term ($\omega^4$) as well. These additional degrees of freedom tend to push the phase transition to higher densities and decrease the size of the jump in density associated with the phase transition. The $\Delta$'s generally do not appear until just before the phase transition and have little effect on the position or size of the phase transition while just replacing the hyperons at these densities. The $\omega\rho$ term has a softening effect at low density, while the $\omega^4$ term has a stiffening effect at high density. Strange quarks never appear in large amounts and not until after the onset of the phase transition.
Stable solutions to the TOV equations (including a crust) are shown in Fig 3 for our hybrid and also equivalent pure hadronic EoS's. We find that the additional interaction terms push the maximum mass closer to that of the hadronic models and allow for hybrid stars $>2$ M$_{\odot}$, as suggested by observations \cite{Fonseca:2021wxt}. Additionally, regardless of composition, the $\omega\rho$ term decreases the tidal deformability and stellar radii, while the $\omega^4$ term increases maximum mass. We also find significant differences in the size of the hybrid branch of the neutron-star families. This happens because the maximum-mass stars have a central density near the phase transition.

\begin{figure*}[t!]
  \includegraphics[trim={0 .3cm 0 .1cm},clip,width=1.01\linewidth]{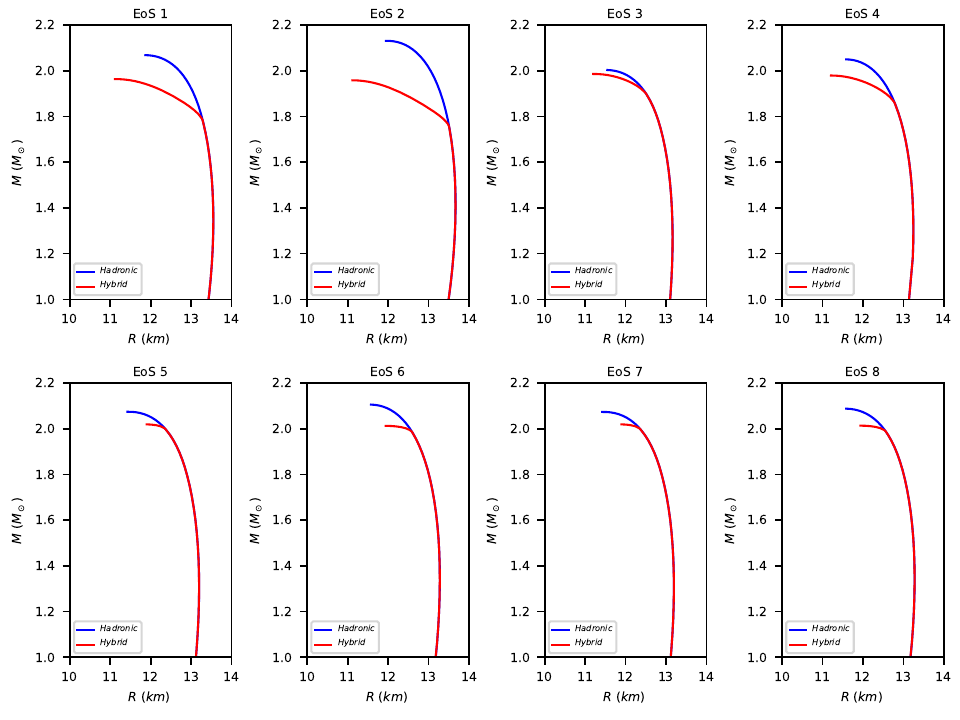}
  \caption{Mass-radius diagram for different EoS's showing only stable stars.\vspace{-.8cm}}
  \label{fig:3}
\end{figure*}


We find that net isospin, strangeness, and weak chemical equilibrium with leptons can considerably change the effects of temperature and magnetic fields on particle content and deconfinement in dense matter. This could be important for detecting quark matter e.g. in neutron star mergers. Although some of our CMF results concerning the magnetic field do not agree with PNJL results, it must be accounted that the PNJL model does not describe hadronic degrees of freedom, and in particular there are no change of degrees of freedom (from nucleonic to hyperonic) at intermediate densities, which is a very important component of the CMF formalism. In the case of a weaker phase transition, we found that adding various degrees of freedom and additional interaction terms to the CMF model can change the onset and size of phase transition, as well as the softness and stiffness in specific density regimes of an EoS. This in turn has significant effects on macroscopic properties of neutron stars. We can reproduce $2$ M$_{\odot}$-stars containing hyperons and $\Delta$'s, as well as deconfined quarks. 
In our future work, we look forward to studying these effects at finite temperatures and in binary neutron-star merger simulations. The zero-temperature hybrid equations of state, together with corresponding hadronic ones, are available on the CompOSE repository  
\cite{Oertel:2016bki,Typel:2013rza,compose} and can be used for different astrophysical applications.

\bibliography{bibliography}

\begin{thebibliography}{17}

\bibitem{Papazoglou:1998vr}
P.~Papazoglou, D.~Zschiesche, S.~Schramm, J.~Schaffner-Bielich, H.~Stoecker,
  W.~Greiner, Phys. Rev. C \textbf{59}, 411 (1999)

\bibitem{Dexheimer:2009hi}
V.A. Dexheimer, S.~Schramm, Phys. Rev. C \textbf{81}, 045201 (2010)

\bibitem{MUSES:2023hyz}
R.~Kumar et~al. (MUSES) (2023)

\bibitem{Peterson:2023bmr}
J.~Peterson, P.~Costa, R.~Kumar, V.~Dexheimer, R.~Negreiros, C.~Providencia,
  Phys. Rev. D \textbf{108}, 063011 (2023)

\bibitem{Meisinger:1995ih}
P.N. Meisinger, M.C. Ogilvie, Phys. Lett. B \textbf{379}, 163 (1996)

\bibitem{Fukushima:2003fw}
K.~Fukushima, Phys. Lett. B \textbf{591}, 277 (2004)

\bibitem{Clevinger:2022xzl}
A.~Clevinger, J.~Corkish, K.~Aryal, V.~Dexheimer, Eur. Phys. J. A \textbf{58},
  96 (2022)

\bibitem{Riley:2019yda}
T.E. Riley et~al., Astrophys. J. Lett. \textbf{887}, L21 (2019)

\bibitem{Miller:2019cac}
M.C. Miller et~al., Astrophys. J. Lett. \textbf{887}, L24 (2019)

\bibitem{LIGOScientific:2018cki}
B.P. Abbott et~al. (LIGO Scientific, Virgo), Phys. Rev. Lett. \textbf{121},
  161101 (2018)

\bibitem{Fonseca:2021wxt}
E.~Fonseca et~al., Astrophys. J. Lett. \textbf{915}, L12 (2021)

\bibitem{Ferreira:2013tba}
M.~Ferreira, P.~Costa, D.P. Menezes, C.~Provid\^encia, N.~Scoccola, Phys. Rev.
  D \textbf{89}, 016002 (2014), [Addendum: Phys.Rev.D 89, 019902 (2014)]

\bibitem{Ferreira:2014kpa}
M.~Ferreira, P.~Costa, O.~Louren\c{c}o, T.~Frederico, C.~Provid\^encia, Phys.
  Rev. D \textbf{89}, 116011 (2014)

\bibitem{Costa:2015bza}
P.~Costa, M.~Ferreira, D.P. Menezes, J.a. Moreira, C.~Provid\^encia, Phys. Rev.
  D \textbf{92}, 036012 (2015)

\bibitem{Oertel:2016bki}
M.~Oertel, M.~Hempel, T.~Kl\"ahn, S.~Typel, Rev. Mod. Phys. \textbf{89}, 015007
  (2017)

\bibitem{Typel:2013rza}
S.~Typel, M.~Oertel, T.~Kl\"ahn, Phys. Part. Nucl. \textbf{46}, 633 (2015)

\bibitem{compose}
\urlstyle{tt}\url{https://compose.obspm.fr/}

\end{thebibliography}

\end{document}